\documentclass[lettersize,journal]{IEEEtran}
\usepackage{amsmath,amsfonts}
\usepackage{algorithmic}
\usepackage{algorithm}
\usepackage{array}
\usepackage[caption=false,font=scriptsize,labelfont=rm,textfont=rm]{subfig}
\usepackage{textcomp}
\usepackage{stfloats}
\usepackage{url}
\usepackage{verbatim}
\usepackage{graphicx}
\usepackage{cite}
\usepackage{bm}
\usepackage{CJKutf8}
\usepackage{color}
\usepackage{amssymb}
\begin{document}
\begin{CJK}{UTF8}{gbsn}	
\title{Linear Model of RIS-Aided High-Mobility Communication System}

\author{
Shuaijun Li, Jie Tang, \textit{Senior Member, IEEE,} Beixiong Zheng, \textit{Senior Member, IEEE,} Xiaokai Song,\\Xiaoqi Qin, \textit{Senior Member, IEEE,} Guixin Pan, and Kai-Kit Wong, \textit{Fellow, IEEE} 
\thanks{This work was supported in part by the National Key Research and Development Program of China under Grant 2024YFE0107900, in part by the National Natural Science Foundation of China under Grant 62222105, and Grant 62201214; in part by the Natural Science Foundation of Guangdong Province under Grant 2024A1515010235, and Grant 2023A1515011753; in part by the Applied Basic Research Funds of Guangzhou under Grant 2024A04J5146; in part by the Fundamental Research Funds for the Central Universities under Grant 2024ZYGXZR087; and in part by the GJYC program of Guangzhou under Grant 2024D03J0006. An earlier version of this paper was presented in part at the 2024 IEEE 100th Vehicular Technology Conference (VTC2024-Fall), Washington, DC, USA, 2024\cite{ref0}. \emph{(Corresponding authors: Jie Tang.)}}
\thanks{Shuaijun Li and Jie Tang are with the School of Electronic and Information Engineering, South China University of Technology, Guangzhou 510640, China (e-mail: eeshuaijunli@mail.scut.edu.cn; eejtang@scut.edu.cn).}
\thanks{Beixiong Zheng is with the School of Microelectronics, South China University of Technology, Guangzhou 511442, China (e-mail: bxzheng@scut.edu.cn).}
\thanks{Xiaokai Song is with the School of Electronics and Information Engineering, Harbin Institute of Technology, Harbin 150001, China (e-mail: songxiaokai@hit.edu.cn).}
\thanks{Xiaoqi Qin is with the State Key Laboratory of Networking and Switching Technology, Beijing University of Posts and Telecommunications, Beijing 100876, China (e-mail: xiaoqiqin@bupt.edu.cn).}
\thanks{Guixin Pan is with the China Unicom Guangdong Branch, Guangzhou 510700, China (e-mail: pangx6@chinaunicom.cn).}
\thanks{Kai-Kit Wong is with the Department of Electronic and Electrical Engineering, University College London, WC1E 7JE London, U.K. (e-mail: kai-kit.wong@ucl.ac.uk).}
}    
\maketitle

\begin{abstract}
Reconfigurable intelligent surface (RIS)-aided vehicle-to-everything (V2X) communication has emerged as a crucial solution for providing reliable data services to vehicles on the road.
However, in delay-sensitive or high-mobility communications, the rapid movement of vehicles can lead to random scattering in the environment and time-selective fading in the channel. In view of this, we investigate in this paper an innovative linear model with low-complexity transmitter signal design and receiver detection methods, which boost stability in fast-fading environments and reduce channel training overhead. Specifically, considering the differences in hardware design and signal processing at the receiving end between uplink and downlink communication systems, distinct solutions are proposed. Accordingly, we first integrate the Rician channel introduced by the RIS with the corresponding signal processing algorithms to model the RIS-aided downlink communication system as a Doppler-robust linear model. Inspired by this property, we design a precoding scheme based on the linear model to reduce the complexity of precoding. Then, by leveraging the linear model and the large-scale antenna array at the base station (BS) side, we improve the linear model for the uplink communication system and derive its asymptotic performance in closed-form. Simulation results demonstrate the performance advantages of the proposed RIS-aided high-mobility communication system compared to other benchmark schemes.
\end{abstract}


\section{Introduction}
\IEEEPARstart{I}{n} June 2023, the IMT-2030 promotion group established six typical application scenarios for the sixth-generation (6G) communication. For 6G wireless network, stringent standards require communication delays ranging from 0.1 to 1 $\rm ms$ and robust transmission capabilities in high-mobility scenarios with velocities between 500 and 1000 $\rm km/h$\cite{ref1}\cite{ref2}\cite{ref22}. In addition, to ensure strict vehicle-to-everything (V2X) communication quality-of-service (QoS) requirements, there is a growing interest in developing intelligent and reconfigurable radio environments, instead of relying solely on traditional transceiver technologies that can only adapt to random wireless channels\cite{ref4}\cite{ref08}\cite{ref01}. However, existing fifth-generation (5G) communication network cannot fully meet the demands of upcoming intelligent transportation scenarios or achieve smart and reconfigurable wireless propagation environments\cite{ref6}. To overcome these challenges, various wireless technologies, including adaptive beamforming, dynamic resource allocation, and next-generation multiple access techniques, have been studied to address these challenges and compensate for deep channel fading\cite{ref7}. However, these technologies are primarily used in transceivers, they cannot reshape the wireless propagation channel through passive signal reflection\cite{ref8}. Furthermore, the unpredictable and dynamic nature of wireless channels in fast-fading environments is still a major constraint on the development of V2X communication. 

Currently, reconfigurable intelligent surface (RIS) has emerged as a cost-effective solution to provide various valuable and appealing functions for next-generation wireless communications and significantly enhance communication performance\cite{ref5}\cite{ref02}\cite{ref03}\cite{ref012}. Specifically, RIS consists of a large number of low-cost passive reflecting elements, each capable of independently adjusting the amplitude or phase of the incident electromagnetic wave. By smartly adjusting many reflecting elements, RIS has the ability to flexibly reshape the wireless propagation channel, enhance signal coverage, improve transmission quality\cite{ref11}\cite{ref12}\cite{ref13}, among others. In addition, RIS can operate in a full-duplex mode through passive reflection, thereby reducing the demand for power-hungry components such as radio frequency (RF) chains and power amplifiers\cite{ref14}. Further, the low profile, lightweight, and conformal geometry of RIS also facilitate its application and development in next-generation wireless communication network\cite{ref22}\cite{ref5}. Besides, integrating RIS with existing network infrastructures such as cellular communications and WiFi systems to enable emerging applications has gradually sparked research interest\cite{ref16}. However, the lack of signal processing capabilities in RIS, combined with numerous passive reflecting elements, can lead to relatively high channel estimation overhead in delay-sensitive, high-mobility scenarios, resulting in longer delays before data transmission\cite{ref13}\cite{ref17}.

To achieve high QoS in RIS-aided V2X communication systems, it is crucial to reduce channel training overhead in fast-fading environments, although practical implementation faces significant challenges\cite{ref19}\cite{ref20}\cite{ref21}. Specifically, in high-mobility scenarios, the signal undergoes rapid, time-varying phase-shifts due to changes in surrounding scatterers and high vehicle speeds, resulting in time-selective fading caused by multipath superposition at the receiver, which significantly degrades communication throughput and reliability. As the user moves at high-speed, the line-of-sight (LoS) component of the channel primarily undergoes phase-shift caused by Doppler. Concurrently, the non-line-of-sight (NLoS) component also exhibits variations in both amplitude and phase across blocks, attributable to random multipath scattering in the environment\cite{ref22}\cite{ref5}. Thus, it is necessary to perform frequent channel estimation within short channel coherence intervals. Moreover, the large-scale array of RIS lacks signal processing capabilities, and its channel estimation overhead is generally proportional to the number of reflecting elements\cite{ref13}\cite{ref17}. As such, estimating the channel from the RIS to the base station (BS) or user is challenging. In practical applications, the cascaded BS-RIS-user channel is usually estimated using transmitted pilot symbols\cite{ref24}. Additionally, several novel estimation approaches have emerged to alleviate channel estimation overhead, including RIS grouping, channel sparsity-based estimation, and reference user-based methods\cite{ref25}\cite{ref26}\cite{ref27}, among others. However, these techniques are generally applicable to low-mobility scenarios with longer channel coherence times and may not fully meet the robustness requirements for high-mobility transmissions.

The intricate features of RIS-aided V2X communication have spurred growing interest in studying reliable transmission in high-mobility scenarios. Early studies on RIS-aided high-mobility communications have primarily focused on the deployment of multiple RISs along the roadside to improve transmission quality between the BS and high-speed users traveling along the road\cite{ref28}. In such scenarios, an innovative RIS-aided downlink communication system was proposed to enable simultaneous transmit diversity and passive beamforming, thereby eliminating the need for CSI from high-mobility users \cite{ref29}. Besides, an alternative and practical approach to mitigating product-distance path loss is the deployment of vehicle-side RIS. This RIS can operate in reflection mode when installed within the vehicle or in refraction mode when applied as a surface coating on the vehicle\cite{ref30}. Additionally, in high-speed train communication, a passive RIS integrated with an active relay is installed on the exterior of the train to facilitate signal transmission from the BS, reducing high carriage penetration loss and increasing the likelihood of establishing LoS channels with the serving BS\cite{ref04}. Interestingly, a novel RIS-aided electromagnetic stealth system was proposed, in which an RIS is installed on the target to intelligently and adaptively reduce the radar detection probability\cite{ref05}. While some studies have explored the application of RIS for integration with current communication infrastructure, its potential to ensure robust transmission in high-mobility scenarios remains largely underexplored. 

On the other hand, designing appropriate signal waveforms and corresponding reception detection algorithms can further enhance communication quality\cite{ref31}\cite{ref32}\cite{ref33}\cite{ref34}. In\cite{ref31}, the author proposed a new transmitter signal design that is nearly optimal relative to the upper bound on the system's average error probability and, by relying solely on channel statistics, reduces channel estimation overhead. In\cite{ref32}, the author presented a multi-user joint constellation design that simultaneously addresses inter-user interference and simplifies the complexity of channel estimation. To reduce dependence on CSI in fast-fading environments, an artificial intelligence model for joint signal design and detection was developed to enable accurate signal detection without requiring channel estimation\cite{ref33}. However, these signal designs are not suitable for the fast-fading Rician channel environment introduced by RIS, and therefore fail to ensure reliable transmission in RIS-aided V2X communication. Besides, in high-mobility environments, employing a simple and robust linear model can help to simplify system analysis, optimize signal processing, and provide reliable performance assessment. However, with the introduction of linear models, the presence of cross-interference during the modeling process can result in rank deficiency issues in the channel\cite{ref35}\cite{ref36}. In view of the limitations of traditional signal design and linear modeling, it is important to develop innovative solutions for RIS-aided V2X communication that enhance high-mobility transmission reliability while employing low-complexity channel estimation methods.

\begin{figure}[!t]
	\centering
	\includegraphics[width=3.5in]{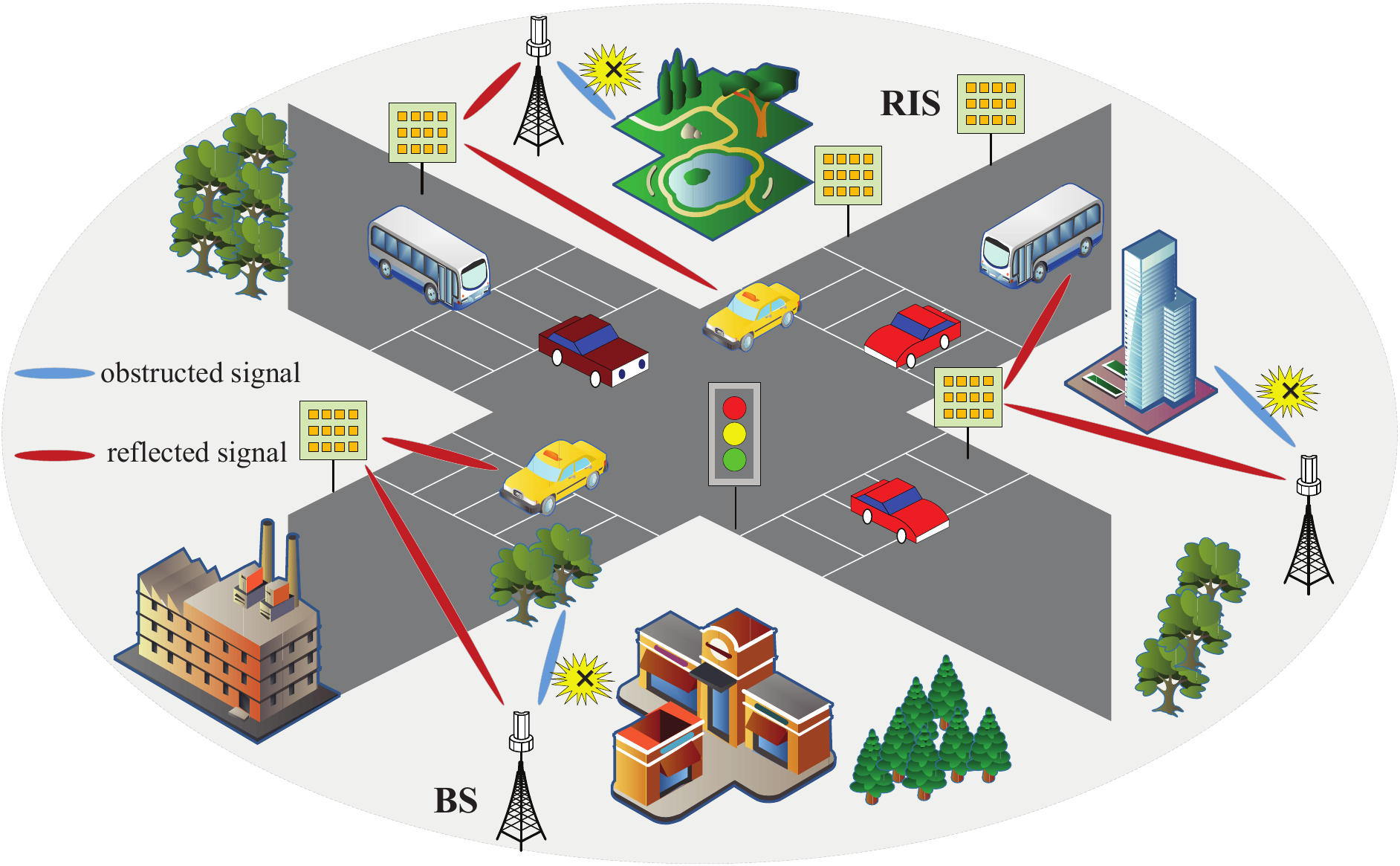}
	\caption{RIS-aided communication for high-speed vehicles.}
	\label{Fig1}
\end{figure}

Inspired by the considerations above, this paper investigates a novel RIS-aided high-mobility communication system\footnote{In this paper, we initially consider an ideal RIS reflection model to simplify the system design. However, practical implementations of RIS are subject to various hardware constraints and imperfections that can limit its signal reflection capabilities, including discrete phase-shifts, phase-dependent amplitude variations, and mutual coupling among the reflecting elements. These challenges will be investigated in future work\cite{ref5}\cite{ref15}.}, as illustrated in Fig. \ref{Fig1}. In this system, multiple multi-antenna BSs, assisted by roadside RISs, achieve reliable transmission and reduce the channel estimation complexity for high-mobility users with partial CSI. Specifically, we consider a new linear model for the downlink communication system, integrating the RIS as an integral part of the BS. Based on this architecture, we introduce a novel linear precoding scheme to mitigate interference among users, without requiring full CSI. Under the above setup, we develop an improved communication model for high-mobility uplink communication systems, along with a closed-form solution to evaluate system performance. The main contributions of this paper are summarized as follows:
\begin{itemize}
\item{First, we propose an efficient scheme to address the challenge of reliable communication in high-mobility environments for RIS-aided downlink communication system. Specifically, deploying multiple RISs along roadsides can establish a strong virtual LoS path between the BS and high-speed users through adjustable signal reflections, transforming multipath fast-fading channels (e.g., Rayleigh fading) into LoS-dominant slow-fading channels (e.g., Rician fading). Attentive to this, we collaboratively design a signal and a matched detection algorithm to transform the downlink communication system into a linear model that suppresses interference.
Subsequently, leveraging the linear model, we design a low-complexity linear precoding scheme to optimize beamforming configuration\footnote{In practice, optimal RIS reflection coefficients can be precalculated offline and stored in the RIS controller. By utilizing a precalculated database that maps AoA information to the corresponding coefficients, RIS can dynamically adjust the signal reflection in real-time. Besides, the reflection coefficient of each RIS element can be controlled using positive-intrinsic-negative (PIN) diodes, which switch between "On" and "Off" states at up to 5 MHz, achieving the desired phase-shift with a rapid switching time of 0.2 ms (e.g.,\cite{ref06}\cite{ref07}).} and mitigate interference among users.}

\item{Next, we present an improved scheme by considering the number of antennas at the BS for RIS-aided high-mobility uplink communication systems, which ensures robust transmission and simplifies channel estimation. Specifically, the uplink system structure is derived from the downlink linear model, enabling the application of the law of large numbers to simplify system analysis. Moreover, to better understand and evaluate system performance, we employ the Gaussian approximation to fit the probability distribution of the received signal and derive a closed-form bound for asymptotic performance.} 

\item{Finally, we provide substantial simulation results to validate the performance of our proposed RIS-aided high-mobility communication system. It is shown that the designed solution effectively transforms the RIS-aided V2X communication system into a Doppler-robust linear model in the real domain, yielding superior performance compared to other benchmark schemes. Besides, extensive simulation results further validate the effectiveness of the Gaussian approximation model and the asymptotic performance analysis.}
\end{itemize}

The rest of this paper is organized as follows. Section \uppercase\expandafter{\romannumeral2} introduces signal design and the channel model for the proposed RIS-aided high-mobility communication system. In Section \uppercase\expandafter{\romannumeral3}, we propose a linear model for RIS-aided downlink communication systems and perform a precoding design. In Section \uppercase\expandafter{\romannumeral4}, we design an improved linear model for RIS-aided uplink communication systems and analyze its performance. Simulation results and discussions are presented in Section \uppercase\expandafter{\romannumeral5}. Finally, conclusions are drawn in Section \uppercase\expandafter{\romannumeral6}.
 
Notations: Upper-case and lower-case boldface letters denote matrices and column vectors, respectively. $\mathbb{R}^{x\times y}$ denotes the space of $x\times y$ real-valued matrices,    $\mathbb{C}^{x\times y}$  denotes the space of $x\times y$ complex-valued matrices, $tr(\cdot)$ denotes the trace of a matrix. $\Re{\{\cdot\}}$ denotes the real component of a complex number. $diag(x)$ denotes a square diagonal matrix with the elements of $x$ on the main diagonal. $\otimes$ denotes the Kronecker product. $\bm I_M $ and $\bm 1_M$ denote an identity matrix and an all-one vector with the dimension of $M$, respectively. Superscripts $ \left(\cdot \right)^T $, $\left(\cdot \right)^* $, $\left(\cdot \right)^H $, and $\left(\cdot \right)^\dagger $ stand for the transpose, conjugate, Hermitian transpose, and Moore-Penrose inverse operations, respectively. $\mathbb{E}[\cdot]$, $\rm var(\cdot)$, and $\|{\cdot}\|$ denote the expectation, variance, and Euclidean norm operators, respectively. $(\cdot)!$ and $C_{m+n}^n$ represent the factorial operator and the combination of selecting $n$ elements from $m+n$, respectively. $x\sim \mathcal{CN}(\mu,\sigma^2)$ means random variable that follows a complex Gaussian distribution with mean $\mu$ and covariance $\sigma^2$, and $\sim$ stands for “distributed as”. $x\sim \mathcal{N}(\mu,\sigma^2)$ means random variable follows a real Gaussian distribution. Finally, $\Phi(\cdot)$ represents the cumulative distribution function of Gaussian. $erf(\cdot)$ denotes the Gaussian error functions. For easy access, the main symbols used in this paper are listed in Table \uppercase\expandafter{\romannumeral1} along with their corresponding meanings. 

\section{System Model}
\subsection{Roadside RIS-aided High-Mobility Communication }
We consider a high-mobility vehicular communication system deploying multiple roadside RISs, as depicted in Fig. \ref{Fig1}, which consists of multiple BSs, roadside RISs, and mobile vehicles. Meanwhile, blue beams represent obstructed signals, while red beams indicate signals reflected by the RIS. It is assumed that these signals are reflected under ideal conditions without any energy loss\cite{ref12}. Specifically, mobile users preferentially connect to nearby cell BSs for communication. However, when obstacles block the LoS channel between the BS and the mobile vehicle, we deploy roadside RISs integrated with advanced signal processing algorithm to enhance communication quality. To ensure generality, our focus centers on the interaction of the BS and the mobile vehicle, assisted by RISs deployed alongside the road\footnote{It is noted that to coordinate user-RIS and RIS-BS handoveris required, optical fiber connections can be used to link the RIS controller with the BS, and wireless transmission can be employed to facilitate low-overhead feedback between the RIS controller and the BS. However, if the RIS has sensing capabilities, it can quickly adjust its beam in response to the received signals (e.g., centralized and distributed RIS handover schemes), thereby eliminating the need for CSI feedback\cite{ref08}.}. Under the time-division duplexing (TDD) protocol, we utilize the uplink-downlink channel reciprocity between the BS and its served vehicles to enhance channel utilization\cite{ref37}. In the context of this exposition, it is assumed that BS is equipped with a uniform linear array (ULA) comprising $N_t$ antennas; while the system supports $N_k$ mobile vehicles, each with a single antenna. Additionally, each RIS is equipped with a uniform planar array (UPA) consisting of $N=N_x\times N_y$ reflecting elements. In this paper, we examine block-fading channels associated with mobile users, assuming channel stability within each transmission block but accounting for variations between blocks due to vehicle mobility\cite{ref08}. 
\begin{table*}[t]
	\centering
	\renewcommand\arraystretch{1.5}
	\caption{LIST OF MAIN SYMBOLS AND THEIR PHYSICAL MEANINGS }
	\label{tab1}
	\begin{tabular}{|c|c|}
		\hline
		\textbf{Symbol} & \textbf{Physical meaning}\\
		\hline
		$N_t$, $N_k$, $N$  & The number of BS antennas, mobile users, and reflecting elements for each RIS.\\ 
		\hline
		$\bm{d}_m\in\mathbb{C}^{1\times N_t} $ & The channel vector between the BS and the $m\operatorname{-th}$ user.\\
		\hline
		$\bm{Q}\in\mathbb{C}^{N\times N_t}$ & The channel matrix from the BS to the RIS.\\
		\hline
		$\bm{Q}^{\rm{LoS}}$/$\bm{Q}^{\rm{NLoS}}$ & The LoS/NLoS components of the channel $\bm{Q}$.\\
		\hline
		$\bm{g}_m\in \mathbb{C}^{1\times N}$ & The channel vector from the RIS to the $m\operatorname{-th}$ user.\\ 
		\hline
		$\bm g_m ^{\rm{LoS}}$/$\bm g_m ^{\rm{NLoS}}$ & The LoS/NLoS components of the channel $\bm{g}_m $.\\
		\hline
		$\bm{e}(\theta,N)$/$\bm{a}_{\rm RIS}(\theta,\phi)$ & The array response vector in the ULA/UPA model.\\
		\hline
		$\bm \Omega\in\mathbb{C}^{N\times N}$ & The RIS reflection matrix.\\
		\hline
		$\bm{H}\in\mathbb{C}^{N_k \times N_t}$ & The cascaded BS-RIS-user channel matrix.\\
		\hline
		$\bm h_m =\bm g_m \bm\Omega \bm{Q}$ & The $m\operatorname{-th}$ row vector of $\bm{H}$.\\
		\hline 
		${h_{mn}}{e^{j{\theta _{mn}}}}$ & The $h_{mn}$ and $\theta_{mn}$ denote the channel magnitude and phase coefficients.\\
		\hline 
		$e^{j{\nu(t)}}$ & The $\nu(t)$ represents the phase rotation induced by the Doppler shift.\\
		\hline 
		$\bm s$, $\bm{\bar s} $ & The amplitude-modulated signal vector and its complementary pair.\\
		\hline 
		$\bm{\bar{H}}\in\mathbb{R}^{N_k \times N_t } $ & The equivalent channel matrix in the linear model.\\
		\hline 
		$\bm{\bar{h}}_m=\Re(\bm h_m^* \bm 1_{N_t} \cdot \bm h_m)$ & The $m\operatorname{-th}$ row vector of $\bm{\bar{H}}$.\\
		\hline 
		$ \bm{\bar{x}}= 2{\bm s} - 1 $ & The equivalent input signal vector.\\
		\hline
		$\bm{P}$ & The precoding coefficient matrix.\\
		\hline
		$\bm {B}=\bm{\bar{H}}\bm{P}$ & The cascade channel matrix after precoding.\\
		\hline
		$\bm{\bar{H}}_p$ & The equivalent channel matrix after precoding.\\
		\hline
		$\bm G \in \mathbb{C}^{N\times{N_k}}$ & The channel matrix from the user to the RIS. \\
		\hline
		$\bm{q}_m\in \mathbb{C}^{1\times N}$ &  The channel vector from the RIS to the $m\operatorname{-th}$ antenna at the BS.\\
		\hline
		$\bm{c}_m =\bm{q}_m \bm \Omega \bm G $ & The cascaded channel vector from the user to the BS via the RIS.\\
		\hline
	\end{tabular}	
\end{table*}

\subsection{Channel Model}
Let $\bm{d}_m\in\mathbb{C}^{1\times N_t} $ denote the channel for the direct link from the BS to the $m\operatorname{-th}$ user, excluding any reflections from RIS. It is considered to remain static within each block but may vary between blocks as a result of user mobility. Considering the fixed positions of the RIS, the channel between the BS and the RIS can be represented as $\bm{Q} \in\mathbb{C}^{N\times N_t}$, is dominated by the LoS path and remains constant throughout the signal frame, with the Rician factor also remaining unchanged. Meanwhile, we represent the channel from the RIS to the $m\operatorname{-th}$ mobile user with $\bm{g}_m\in \mathbb{C}^{1\times N} $, which also exhibits LoS characteristics. Hence, for large propagation distances, we model the link as a Rician channel based on a far-field LoS plane wavefront. To simplify the analysis, we assume the vehicle moves at a constant speed in an open area with a limited number of scatterers, resulting in a constant Doppler shift within each time block (without Doppler jitter), while neglecting the direct path\cite{ref08}. The BS-RIS and RIS-user ($m\operatorname{-th}$) channels can be respectively modeled as
\begin{subequations}\label{eqn1}
	\begin{align}
		\bm{Q} &=\sqrt{\frac{K}{1+K}}\bm{Q}^{\rm{LoS}}+\sqrt{\frac{1}{1+K}}\bm{Q}^{\rm{NLoS}}\label{deqn1A},\\
		\bm g_m &=\sqrt{\frac{V}{1+V}} \bm g_m ^{\rm{LoS}}+\sqrt{\frac{1}{1+V}}\bm g_m ^{\rm{NLoS}}\label{deqn1B},	
	\end{align}
\end{subequations}
where $K$ and $V$ denote the Rician factors. $\bm{Q}^{\rm{LoS}}$ and $\bm g_m ^{\rm{LoS}}$ represent the LoS component of the channel; while $\bm{Q}^{\rm{NLoS}}$ and $\bm g_m ^{\rm{NLoS}}$ denote the NLoS component, and follow the complex Gaussian distribution. In the ULA model used at the BS, the array response vector can be denoted by $\bm{e}(\theta,N_t)=[1, \cdots, e^{j2\pi(N_t-1)\frac{d}{\lambda}\sin\theta}]$, with $\theta$ denoting the angle-of-arrival (AoA) or angle-of-departure (AoD), $N_t$ indicates the antenna amount in the one-dimensional (1D) array, $d$ represents the distance among antenna elements and $\lambda $ corresponds to the signal wavelength. Moreover, for the UPA model at the RIS, the array response vector is expressed as the Kronecker product of two 1D steering vector functions\cite{ref28}\cite{ref30}, i.e., 
\begin{flalign}\label{eqn2}
	\begin{split}
		&\bm{a}_{\rm RIS}(\theta,\phi)=\bm{e}(\theta,N_x) \otimes \bm{e}(\theta,N_y)\\
		&=\sqrt{\frac{1}{N_x N_y}} [1, \cdots ,e^{j\frac{2\pi}{\lambda}d(m\sin\phi\sin\theta+n\cos\theta)}, \cdots],
	\end{split}   
\end{flalign}
where $\theta\in [0,\pi]$ and $\phi\in [0,2\pi)$ represent the elevation and the azimuth angle of AoA or AoD; while $N_x$ and $N_y$ represent the numbers of antennas in the horizontal and vertical directions, respectively, with $m\in[0,\cdots ,N_x-1]$ and $n\in[0,\cdots ,N_y-1]$. Based on the discussions above, the LoS component in the Rician channel is represented by the product of the steering vector, we have  
\begin{subequations}\label{eqn3}
	\begin{align}
		\bm{Q}^{\rm LoS} &=\bm{a}_{\rm RIS}(\theta,\phi)\bm{a}_{\rm BS}^H (\theta,\phi)\label{deqn3A},\\
		\bm{g}^{\rm LoS} &=\bm{a}_{\rm UE}(\theta,\phi)\bm{a}_{\rm RIS}^H (\theta,\phi)\label{deqn3B}.	
	\end{align}
\end{subequations}
Additionally, let $\bm \Omega=diag(e^{j\varphi_1},\cdots,e^{j\varphi_N})\in\mathbb{C}^{N\times N}$ denotes the RIS reflection matrix, where $\varphi_n$ represents the reflection phase-shift of $n\operatorname{-th}$ element. Further, the cascaded BS-RIS-user ($m\operatorname{-th}$) channel can be expressed as  
 \begin{flalign}\label{eqn4}
	\begin{split}
		\bm h_m &=\bm g_m \bm\Omega \bm{Q}\\
		=&\sqrt{\frac{V}{1+V}}\sqrt{\frac{K}{1+K}}\bm{g}_m^{\rm{LoS}}\bm\Omega \bm{Q}^{\rm{LoS}}\\
		&+\sqrt{\frac{V}{1+V}}\sqrt{\frac{1}{1+K}}\bm{g}_m^{\rm{LoS}}\bm\Omega \bm{Q}^{\rm{NLoS}}\\
		&+\sqrt{\frac{1}{1+V}}\sqrt{\frac{K}{1+K}}\bm{g}_m^{\rm{NLoS}}\bm\Omega \bm{Q}^{\rm{LoS}}\\
		&+\sqrt{\frac{1}{1+V}}\sqrt{\frac{1}{1+K}}\bm{g}_m^{\rm{NLoS}}\bm\Omega \bm{Q}^{\rm{NLoS}}.
	\end{split}   
\end{flalign}
\subsection{Signal Design}
In this section, we utilize RIS capabilities to generate virtual direct LoS paths and create Rician fading scenarios. By designing an orthogonal, dual-frequency, and multi-amplitude complementary signal\footnote{Note that the system model and signal design in this paper differ from those in parallel works\cite{ref36}\cite{ref09}, which also consider a linear model based on channel magnitude estimation but experience cross-product interference across spatial channels, leading to a rank deficiency problem that, in turn, results in an expanded spatial dimension of the system model and increases detection complexity.}, subtraction operations are employed to construct opposing cross-product terms, thereby eliminating interference during the linear modeling process and reducing channel training overhead, i.e.
\begin{equation}\label{eqn5}
	x_n(t) = \frac{1}{{A - 1}}({s_n}{e^{j{w_1}t}} + {\bar s_n}{e^{j{w_2}t}}),
     \end{equation}
where ${w_n} = 2\pi {f_n}$, the frequencies ${f_1}$ and ${f_2}$ are orthogonal to each other, $f_{\rm 2}-f_{\rm 1}= 1/T_s $, $T_s$ is symbol duration. In addition, ${s_n} \in [0,1, \cdots ,A - 1]$ is amplitude modulation (AM) signal with $A$ levels on the $n\operatorname{-th}$ transmitting antenna, and together with ${\bar s_n} = A - 1 - {s_n}$ forms a complementary signal such that the sum of their amplitudes is $A - 1$. For simplicity, we take $A = 2$ as an example in our analysis and omit the time index. The signal received at the $m\operatorname{-th}$ antenna can be expressed as 
\begin{equation}\label{eqn6}
	y_m  = \bm h_m \bm x + n_m, 
\end{equation}
where $\bm h_m = [{h_{m1}}{e^{j{\theta _{m1}}}}, \cdots ,{h_{mN_t}}{e^{j{\theta _{mN_t}}}}]$  indicates the $m\operatorname{-th}$ row of $\bm{H}\in\mathbb{C}^{N_k \times N_t}$, $h_{mn}$ and $\theta_{mn}$ denote the channel's magnitude and phase, $\bm x$  is a transmit signal vector, and $n_m \sim \mathcal{CN} (0,{\sigma ^2})$ represents the additive white Gaussian noise (AWGN). Additionally, symbol vectors are written as $ \bm s = [{s_1}, \cdots {s_{N_t}}]^T \in \Im $ and $\bm{\bar s} = 1 -\bm s $, where $\Im $ represents the constellation set for transmitting vectors. 

The received signal $y_m$, after processing by the correlator, is expressed as 
\begin{equation}\label{eqn7_1}
	y_m^{(k)} = \frac{1}{T_s}\int_{0}^{T_s} y_m e^{-j{w_k}t} dt, 
\end{equation}
when ${w_n} \ne {w_k} $, and the ${w_n}$ and ${w_k}$ are orthogonal to each other, there is $| y_m^{(k)}|\approx 0 $. Conversely, when ${w_n} = {w_k} $, we have $| y_m^{(k)}|= 1 $.

\section{Linear Model for RIS-aided Downlink Communication System}
As mentioned in Section \uppercase\expandafter{\romannumeral2-A}, when vehicles travel at high-speed on roads with obstacles obstructing direct communication to the BSs, we can introduce RIS to enhance the capability of communication. However, high-speed movement can introduce interference that significantly impacts reliable communication. Besides, achieving real-time Doppler compensation and obtaining instantaneous CSI in fast-fading scenarios also present certain challenges in practice. As such, in this section, we design a scheme for RIS-aided downlink communication in high-mobility scenarios. Under the premise of sacrificing certain performance, transforming the system into a linear model in the real domain aims to enhance the system's Doppler robustness.

\subsection{Linear Model Design}
For the $m\operatorname{-th}$ mobile user, the received signal ${y_m}$, after being processed by the correlator at  ${f_1}$ and ${f_2}$, respectively, is given by
\begin{subequations}\label{eqn7}
	\begin{align}
		y_m^{(1)} &= \frac{1}{T_s}\int_{0}^{T_s} e^{j{\nu(t)}} y_m e^{-j{w_1}t} dt  \nonumber \\
		&=\sum\limits_{n = 1}^{{N_t}} e^{j{\nu(t)}} h_{mn} {e^{j{\theta _{mn}}}}{s_n} + n_m^{(1)},\\
		y_m^{(2)} &= \frac{1}{T_s}\int_{0}^{T_s} e^{j{\nu(t)}} y_m e^{-j{w_2}t} dt  \nonumber \\
		&=\sum\limits_{n = 1}^{{N_t}} e^{j{\nu(t)}} h_{mn} {e^{j{\theta _{mn}}}}{\bar s_n} + n_m^{(2)},	
	\end{align}
\end{subequations}
where $\nu(t)=2\pi f_d(t)$ represents the phase rotation induced by the Doppler shift $f_d$. However, to simplify the analysis, $\nu(t)$ is typically treated as a constant $\nu$ within a symbol period \footnote{In far-field environments, where the propagation distance significantly exceeds the antenna array size, all parallel paths are expected to experience nearly identical Doppler shifts. Although spatial paths may experience varying Doppler shifts due to different AoA, their impact on the channel is minimal, especially in scenarios with a strong LoS component\cite{ref010}.}. Due to the orthogonality of ${f_1}$ and ${f_2}$, $n_m^{(1)}$ and $n_m^{(2)}$ remain an independent and identically distributed $(i.i.d.)$ complex Gaussian distribution after processing by the correlator. Subsequently, the receiver implements a low-complexity signal detection, we have  
\begin{equation}\label{eqn8}
	\begin{split}
		z_m &=\left|y_m^{(1)} \right|^2 -\left|y_m^{(2)} \right|^2\\
		&=\sum\limits_{n = 1}^{N_t} \sum\limits_{k = 1}^{N_t}h_{mn}h_{mk}{e^{j(\theta_{mn} - \theta_{mk})}}{s_n}{s_k}\\
		&-\sum\limits_{n = 1}^{N_t} \sum\limits_{k = 1}^{N_t}h_{mn}h_{mk}{e^{j(\theta_{mn} - \theta_{mk})}}{\bar s_n}{\bar s_k}+ u_m\\
		&=\sum\limits_{n = 1}^{N_t} \sum\limits_{k = 1}^{N_t}h_{mn}h_{mk} \cos{(\theta_{mn} - \theta_{mk})}{(2{s_n} - 1)}+ u_m\\
		&= \Re{(\bm h_m ^ \ast \bm 1_{N_t} \cdot \bm h_m)}\bm{\bar{x}}+ u_m,
	\end{split}   
\end{equation}
where $\bm{\bar{x}}= 2{\bm s} - 1 $ represents the equivalent input signal vector, which can be regarded as a bipolar amplitude mapping of source symbols; while $u_m$ denotes the equivalent noise. 

After processing at the receiving end, the system is linearly modeled in the real domain, and the Doppler phase rotation $e^{j\nu}$ caused by fast-fading environments is also eliminated, we obtain
\begin{equation}\label{eqn9}
	z_m =\bm{\bar{h}}_m \bm{\bar{x}} + u_m.	
\end{equation}
The $u_m$ can be expressed as 
\begin{equation}\label{eqn10}
	u_m = 2\Re(\bm h_m \bm s {n_m^{(1)}}^*) + {\left|n_m^{(1)} \right|^2} - 2\Re(\bm h_m  \bm {\bar s} {n_m^{(2)}}^*) -{\left|n_m^{(2)} \right|^2}.
\end{equation} 
Moreover, $\bm{\bar{H}}=[\bm{\bar{h}}_1,\cdots,\bm{\bar{h}}_{N_k}]^T\in\mathbb{R}^{N_k \times N_t } $ represents the equivalent channel in the linear model, it can be efficiently estimated using the methods described in \cite{ref36} and \cite{ref011}. $\bm{\bar{h}}_m=\Re(\bm h_m^* \bm 1_{N_t} \cdot \bm h_m) $ denotes the row vector of $\bm{\bar{H}}$, which can be written as  
\begin{equation}\label{eqn11}
	\bm{\bar{H}}=\Re(\bm \Lambda \bm{H}) = \frac{1}{2}[(\bm \Lambda \bm{H}) + {(\bm {\Lambda} \bm{H})^*}],	
\end{equation} 
where $ \bm {\Lambda} $ is a diagonal matrix with its diagonal elements represented by $ \Lambda_{mm} = h_{m1}e^{-j\theta_{m1}} + \cdots  + h_{m{N_t}}e^{-j\theta_{m{N_t}}} $.

Using the linear model outlined above, the receiver can implement joint minimum Euclidean norm detection with multiple antennas, we have
\begin{equation}\label{eqn12}
	\hat{\bm s} = \mathop {\arg \min }\limits_{\bm s \in \Im} \sum\limits_{m = 1}^{{N_k}} \left\Vert {z_m -\bm{\bar{h}}_m \bm{\bar{x}}}\right\Vert^2. 
\end{equation}
From the \eqref{eqn12}, it is evident that using a real-domain linear model for receiver signal detection can help avoid the high channel estimation overhead associated with fast-fading channels, since the $\bm{\bar{h}}_m$ varies relatively slowly in fast-fading environments. However, relying solely on partial amplitude channel information for signal detection enhances robustness at the cost of performance loss. Besides,
in the linear system, a simpler least-square (LS) estimate of $\bm{\bar{h}}_m $ is given by

\begin{equation}\label{eqn13}
	\bm{\hat{\bar{h}}}_m=\bm z_{t}\bm{\bar{x}}^T_{t}(\bm{\bar{x}}_{t}\bm{\bar{x}}^T_{t})^{-1},	
\end{equation}
where $\bm{\bar{x}}_{t}$ represents the equivalent signal of the pilot symbols used for channel estimation, $\bm z_{t}$ denotes the pilot signal received at the receiver.
\subsection{Precoding Design Based on Linear Model}
From the analyses above, it is clear that after nonlinear processing at the receiver, the system can be effectively modeled as a Doppler-robust linear model. To simplify precoding design and reduce channel feedback overhead, the transmitter can use the linear model for precoding. However, since the transmitter utilizes equivalent channel $\bm{\bar{H}}$ for precoding, there is a performance loss compared to the traditional method that employs real-time CSI for precoding. In this approach, $\bm{P}$ represents the precoding coefficient matrix, and $\bm {B}=\bm{\bar{H}}\bm{P}$ denotes the cascade channel matrix after precoding. To further minimize interference among systems, we designe the equivalent channel matrix after precoding to be an identity matrix, i.e.	
\begin{equation}\label{eqn14}
	\bm{\bar{H}}_p=\Re(\bm{\bar{\Lambda}}\bm {B}) = \frac{1}{2}[(\bm{\bar{\Lambda}}\bm {B}) + (\bm{\bar{\Lambda}} \bm {B})^*]=\bm I, 		
\end{equation}
where $\bm{\bar{\Lambda}}$ is also a diagonal matrix with its diagonal elements represented by $ \bar{\Lambda}_{ll} = b_{l1}e^{-j\theta_{l1}} +\cdots + b_{l{N_t}}e^{-j\theta_{l{N_t}}} $, and $b_{ln}e^{j\theta_{ln}}$ denotes the entry in the $l\operatorname{-th}$ row and $n\operatorname{-th}$ column of $\bm {B}$. In addition, since $\bm{\bar{\Lambda}}$ is a diagonal matrix, to design the precoding matrix as an identity matrix, $\bm {B}$ must be diagonal and satisfy
\begin{equation}\label{eqn15}
	\bar{\Lambda}_{ll}b_{ll} e^{j\theta_{ll}} = 1.
\end{equation}	
Given that $\bm {B}$ is diagonal and must meet condition ${b_{lk}}{e^{j{\theta_{lk}}}}\,(l \ne k) = 0$, the \eqref{eqn15} can be further written as
\begin{equation}\label{eqn16}
	\left(b_{ll} e^{-j\theta_{ll}}\right) \cdot b_{ll} e^{j\theta_{ll}}=\left|b_{ll}\right|^2 = 1.	
\end{equation}	
The condition satisfied by $\bm{P}$ can be designed as
\begin{equation}\label{eqn17}
	\left|\bm {B} \right|=\left|\bm{\bar{H}}\bm{P}\right|=\bm I,	
\end{equation}
when only condition $\bm{\bar{H}}\bm{P} = \bm I $ is considered, we use a zero-forcing (ZF) precoding design at the transmitter, and the precoding matrix is given by	
\begin{equation}\label{eqn18}
	\bm{P}=\bm{\bar{H}}^H (\bm{\bar{H}}\bm{\bar{H}}^H)^{-1}=\bm{\bar{H}}^\dagger.
\end{equation}

At the receiver, the low-complexity signal detection method can also be employed, we obtain 
\begin{equation}\label{eqn19}
	\left|\bm{\bar{H}}\bm{P}\cdot\frac{\bm 1_{N_t}+\bm{\bar{x}}}{2}\right|^2-\left|\bm{\bar{H}}\bm{P}\cdot\frac{\bm 1_{N_t}-\bm{\bar{x}}}{2}\right|^2=\bm{\bar{x}}.	
\end{equation}
The \eqref{eqn19} shows that precoding eliminates inter-user interference, enabling the receiver to accurately recover equivalent transmitted signal. Moreover, the output SNR at the receiver is given by
\begin{equation}\label{eqn20}
	\eta =	\frac{\rho\mathbb{E}[s_n^2]}{\mathbb{E}[u_n^2]}= \frac{\rho}{2\sigma^2+3\sigma^4 / 4},
\end{equation}	
where $\rho $ is the amplitude gain of the transmission signal, the overall average transmitted power can be written as
\begin{equation}\label{eqn21}
	\begin{split}
		p_s &= \rho \mathbb{E}[\left|\bm{P} \bm s\right|^2 + \left|\bm{P} \bm{\bar s}\right|^2] \\
		&= 2\rho \cdot tr(R_{\bm{\bar{H}}^\dagger}^{-1}),
	\end{split}   	
\end{equation}
where $R_{\bm{\bar{H}}^\dagger}^{-1}=(\bm{\bar{H}}^\dagger)^H \bm{\bar{H}}^\dagger$ represents the inverse of  correlation matrix of $\bm{\bar{H}}^\dagger$, assuming the overall transmitted power is 1, we have 
\begin{equation}\label{eqn22}
	\rho=1/2tr(R_{\bm{\bar{H}}^\dagger}^{-1}).	
\end{equation}
In high SNR scenarios that meet condition $\sigma^4 \approx 0 $, the resulting output SNR from precoding can be rewritten as
\begin{equation}\label{eqn23}
	\eta \approx \frac{1}{4tr(R_{\bm{\bar{H}}^\dagger}^{-1})\sigma^2} \approx \frac{N_t-N_k-1}{4N_k\sigma^2}.
\end{equation}
\section{Improved Linear Model for RIS-aided Uplink Communication System}
For the RIS-aided downlink communication system proposed in Section \uppercase\expandafter{\romannumeral3}, we propose a Doppler-robust linear model to ensure reliable communication in high-mobility scenarios. Further, in uplink communication systems, the large number of receiving antennas allows for system model simplification using the law of large numbers and facilitates performance analysis through simpler methods. The next two subsections will offer detailed explanations of these aspects.

\subsection{Improved Linear Model Design}
In accordance with the reciprocity principle of the uplink and downlink channels, in the RIS-aided uplink communication system, $\bm G \in \mathbb{C}^{N\times{N_k}}$ and $ \bm{q}_m \in \mathbb{C}^{1\times N}$ respectively represent channels dominated by LoS path from the user to the RIS, and from the RIS to the $m\operatorname{-th}$ antenna at the BS.  We denote the cascaded channel from the user to the BS via the RIS as 
\begin{equation}\label{eqn24}
	\begin{split}
		\bm{c}_m &=\bm{q}_m \bm \Omega \bm G  \\
		&=\left(\sqrt{\frac{V}{1+V}}\bm{q}_m^{\rm LoS}+\sqrt{\frac{1}{1+V}}\bm{q}_m^{\rm NLoS}\right)\bm \Omega\\
	&\left(\sqrt{\frac{K}{1+K}}\bm G^{\rm LoS}+\sqrt{\frac{1}{1+K}} \bm G^{\rm NLoS}\right),
	\end{split}   	
\end{equation}
where $\bm{q}_m^{\rm LoS}$ and $\bm G^{\rm LoS}$ represent the LoS component of the channel, which in the context of Rician fading, are obtained by the product of steering vectors. Meanwhile, $\bm{q}_m^{\rm NLoS}$ and $\bm G^{\rm NLoS}$ denote the NLoS component and follow a zero-mean complex Gaussian distribution. For ease of notation, the above expression can be simplified as $\bm{c}_m =a+b+o$, where
\begin{subequations}\label{eqn25}
	\begin{align}
		a &=\sqrt{\frac{V}{1+V}}\bm{q}_m^{\rm{LoS}} \bm \Omega \sqrt{\frac{K}{1+K}}\bm G^{\rm LoS}, \\
		b &=\sqrt{\frac{V}{1+V}}\bm{q}_m^{\rm{LoS}} \bm \Omega \sqrt{\frac{1}{1+K}}\bm G^{\rm NLoS} \nonumber \\
		&+\sqrt{\frac{1}{1+V}}\bm{q}_m^{\rm{NLoS}}\bm \Omega \sqrt{\frac{K}{1+K}} \bm G^{\rm LoS}, \\
		o &=\sqrt{\frac{1}{1+V}}\bm{q}_m^{\rm{NLoS}}\bm \Omega \sqrt{\frac{1}{1+K}} \bm G^{\rm NLoS}. 
	\end{align}
\end{subequations} 
As described in Section \uppercase\expandafter{\romannumeral2}-C, the signal received at the $m\operatorname{-th}$ antenna of the BS, transmitted through the user-RIS-BS channels, is given by 
\begin{equation}\label{eqn26}
\tilde{y}_m=\bm{c}_m \bm x+v_m,
\end{equation} 
where $ v_m \sim \mathcal{CN} (0,\sigma^2) $ represents the AWGN, $\bm{x}$ denotes complementary signal vector. 
After processing by a correlator and other devices at the receiver, $\tilde{z}_m$ can also be equivalently modeled as a linear model, i.e.
\begin{equation}\label{eqn27}
\begin{split}
		\tilde{z}_m &=\left|\bm{c}_m \bm s + v_m^{(1)} \right|^2-\left|\bm{c}_m \bm{\bar{s}} + v_m^{(2)}\right|^2\\
		             &=\Re(\bm{c}_m^* \bm 1_{N_k} \cdot \bm {c}_m ) \bm{\bar{x}} + {\varsigma}_m,
\end{split}   	
\end{equation}
where $\bm{\bar{x}} = 2 \bm{s}-1$ represents the equivalent input signal vector, $ v_m^{(k)}\sim \mathcal{CN} (0,\sigma_v^2)$ denotes $i.i.d.$ complex Gaussian variable with variance $\sigma_v^2=\sigma^2/2$ after processing by the correlator, and ${\varsigma}_m $ denotes the equivalent noise. After statistically averaging the signals received by each antenna, we have  
\begin{equation}\label{eqn28}
\xi=\frac{1}{N_t}\sum\limits_{m = 1}^{N_t} \tilde{z}_m.  
\end{equation} 

The channel characteristics described in \eqref{eqn25} allow the \eqref{eqn28} to be decomposed into the sum of several components. It can be rewritten as $\xi=\xi_1+\xi_2+\xi_3+\xi_4 $, where $\xi_1 $ can be characterized as  
\begin{equation}\label{eqn29}
\begin{split}
\xi_1 &= \frac{1}{N_t} \sum \limits_{m = 1}^{N_t} \Re(a_m^*{1_{N_k}} \cdot {a_m}) \bm{\bar{x}}\\
              &= \sum \limits_{n = 1}^{N_k} \left[\sum \limits_{j = 1}^{N_k} \frac{\bm 1}{N_t} \sum \limits_{m = 1}^{N_t}\Re(a_{mj}^*a_{mn}) \right]\bar{x}_n\\
              &= \sum\limits_{n = 1}^{N_k} \varrho_n^{(\xi_1)} \bar{x}_n, 
\end{split}   	
\end{equation}
where $\varrho_n^{(\xi_1)}=\sum \limits_{j = 1}^{N_k} \frac{\bm 1}{N_t} \sum \limits_{m = 1}^{N_t}\Re(a_{mj}^*a_{mn})$. Similarly, $\xi_2 $, $\xi_3$, $\xi_4$, $\varrho_n^{(\xi_2)}$, $\varrho_n^{(\xi_3)}$, and $\varrho_n^{(\xi_4)}$,  can also be expressed in a similar form. 

As a result, $\xi$ can be rewritten as
\begin{equation}\label{eqn31}
\xi = \sum \limits_{n = 1}^{N_k} \varrho_n\bar{x}_n,
\end{equation}
where $\varrho_n = \varrho_n^{(\xi_1)}+\varrho_n^{(\xi_2)}+\varrho_n^{(\xi_3)}+\varrho_n^{(\xi_4)}.$ Additionally, by utilizing \eqref{eqn31} and the equivalent transmitted signal $\bar{x}_n = 2 {s_n}-1$, a simplified channel estimation method can be designed, significantly reducing channel training overhead in fast-fading environments. Specifically, in estimating the $\varrho_n $, with the transmitted signal satisfying ${s_n} = 1$, and ${s_m} = 1/2 \;(m \ne n,m \in [1,{N_k}])$, we can obtain $\xi \approx \varrho_n $. 

Based on the above analyses, it is evident that $\varrho_n^{(\xi_1)}$ is composed of the LoS component of the channel, which plays a dominant role in $\varrho_n$. In contrast, when the number of receiving antennas is sufficiently large, $\varrho_n^{(\xi_2)}$ and $\varrho_n^{(\xi_3)}$ exhibit minor fluctuations, mainly governed by the variance of the channel's NLoS component. Furthermore, as the mean of the NLoS component is zero, we have $\varrho_n^{(\xi_4)} \approx 0$. Hence, in a Rician channel environment dominated by the LoS path and with a sufficient number of receiving antennas, we have $\varrho_n^{(\xi_1)} \approx \varrho_n$, further simplifying the channel estimation process in high-mobility scenarios. Meanwhile, employing LoS paths for information transmission also ensures channel stability under mobile conditions, thereby enhancing the system's Doppler robustness.

\subsection{Performance Analysis}
\begin {figure*}[hb]
\hrulefill
\begin{equation} \label{eqn034}\tag{34}
	f_{\tilde{z}_m} \left(x\right)=	
	\begin{cases}
		\exp \left(-{\frac{x+\gamma+\gamma^{\prime}}{\beta}} \right)\sum \limits_{k,m = 0}^{+\infty} \sum \limits_{n = 0}^{k} \frac{1}{(km)!} \frac{\gamma^k {\gamma^{\prime}}^m }{\beta^{k+m}} \frac{C_{m+n}^n x^{k-n}}{2^{1+m+n}\beta^{1+k-n}(k-n)!}, x \geq 0\\
		\exp \left(-{\frac{x-\gamma-\gamma^{\prime}}{\beta}} \right)\sum \limits_{k,m = 0}^{+\infty} \sum \limits_{n = 0}^{m} \frac{1}{(km)!} \frac{\gamma^k {\gamma^{\prime}}^m }{\beta^{k+m}} \frac{C_{k+n}^n (-x)^{m-n}}{2^{1+k+n}\beta^{1+m-n}(m-n)!}, x < 0.	
	\end{cases}
\end{equation}
\end{figure*}
The improved linear model in the uplink system suppresses interference but also leads to performance loss. Thus, it is necessary to analyze its theoretical performance. Importantly, this analysis can be simplified in the uplink system by employing a Gaussian approximation method. Besides, it is not difficult to deduce that $\tilde{y}_m$ follows a complex Gaussian distribution with a non-zero mean. After being processed through the correlator and squared, we have $\varepsilon_m =|\bm{c}_m \bm s + v_m^{(1)} |^2.$ It is a generalized gamma random variable
with parameters $\alpha ,\beta$ and $\gamma$, denotes as $\varepsilon_m \sim G(\alpha ,\beta ,\gamma),$ its probability density function (pdf) is given by
\begin{equation}\label{eqn33}
 f_{\varepsilon_m} (x)
 =\frac{1}{\beta} \left(\frac{x}{\gamma}\right)^\frac{\alpha-1}{2}{\rm I}_{\alpha - 1} \left(\frac{2 \sqrt{\gamma x}} {\beta } \right) \exp \left(- {\frac{\gamma+x}{\beta }} \right), x\geq 0,
\end{equation}
where $\alpha=1$, $\beta=2\sigma_v^2$, and $\gamma=\left|\bm {c}_m \bm{s} \right|^2 $; while $\rm I_n(\cdot)$ represents the first kind modified $n\operatorname{-th}$ order Bessel function. 

{\it Proof:} Please refer to Appendix. $\hfill\blacksquare$ 
 
After nonlinear processing, the distribution at the receiver is characterized by the difference between two generalized gamma distributions, the result is directly taken from \cite[(2)]{ref38}, we obtain
\begin{equation}\label{eqn34}
\tilde{z}_m \sim \left[G(\alpha ,\beta ,\gamma)-G(\alpha^{\prime} ,\beta^{\prime} ,\gamma^{\prime})\right],	
\end{equation}  
where $\alpha^{\prime}= 1$, $\beta^{\prime}=2\sigma_v^2$, and $\gamma^{\prime}=\left|\bm {c}_m \bm{\bar s} \right|^2 $, the pdf of $\tilde{z}_m$ is shown in \eqref{eqn034} at the bottom of the page, it can be directly obtained from \cite[(3) \& (4)]{ref38} by simple operations.

Under conditions of high SNR and a sufficient numbers of antennas, the complex probability distribution in \eqref{eqn034} can be approximated by a Gaussian model characterized by mean and variance \cite{ref39}, we obtain 
\begin{subequations}\label{eqn35}
\begin{align}
\mu_{\tilde{z}_m} &=\left|\bm {c}_m \bm{s} \right|^2-\left|\bm {c}_m \bm{\bar s} \right|^2 \tag{35a},\\
\sigma_{\tilde{z}_m}^2 &=4\sigma_v^2(\left|\bm {c}_m \bm{s} \right|^2+\left|\bm {c}_m \bm{\bar s}
\right|^2 ) \nonumber \\
&+ \text{var} (|v_m^{(1)} |^2)+ \text{var} (|v_m^{(2)}|^2)\tag{35b}.
\end{align}
\end{subequations} 

Since $v_m^{(k)}$ is a complex Gaussian random variable, $|v_m^{(k)}|^2$ clearly follows an exponential distribution with parameter $1/2\sigma_v^2$ and $\text{var}(|v_m^{(k)}|^2)=4\sigma_v^4 $, $\sigma_{\tilde{z}_m}^2$ can be rewritten as 
\begin{equation}\label{eqn37}
\sigma_{\tilde{z}_m}^2=4\sigma_v^2\left(\left|\bm {c}_m\bm{s} \right|^2+\left|\bm {c}_m \bm{\bar s} \right|^2 \right)+8\sigma_v^4 \tag{36}.		
\end{equation}  
Therefore, under high SNR conditions, the distribution of $\tilde{z}_m$ can be approximated as a Gaussian distribution over the real domain, which is denoted by 
\begin{equation}\label{eqn38} \tag{37}
	\tilde{z}_m \sim \mathcal{N}(\mu_{\tilde{z}_m},\sigma_{\tilde{z}_m}^2).	
\end{equation} 

Similarly, the distribution of $\xi$ can also be approximated, is given by
\begin{equation}\label{eqn39} \tag{38}
\xi	\sim \mathcal{N}(\mu_{\xi},\sigma_{\xi}^2),	
\end{equation} 
where $\mu_{\xi}= \sum \limits_{m = 1}^{N_t} \mu_{\tilde{z}_m }/N_t $, $\sigma_{\xi}^2= \sum \limits_{m = 1}^{N_t} \sigma_{\tilde{z}_m}^2/N_t^2 $. Utilizing the statistical characteristics of Gaussian approximation, the receiver can implement maximum likelihood (ML) detection, yielding 
\begin{equation}\label{eqn40}\tag{39}
\hat{\bm s} = \mathop {\arg \min } \limits_{\bm s \in \Im} \left[ \rm ln (\sigma_{\xi}^2)+\left|\xi-\mu_{\xi}\right|^2/\sigma_{\xi}^2 \right]. 	
\end{equation}

Moreover, the preceding analysis indicates that when the transmitted signal is $ \bm s_r$ and the equivalent input signal is $\bm{\bar{x}}^{(r)}$, $\xi$ can be expressed as $\xi^{(r)} = \sum \limits_{n = 1}^{N_k} \varrho_n \bar{x}_n^{(r)}$. With knowledge of channel and noise statistics, we can partition the observation space into multiple decoding regions, denoted as $\left\{d_r\right\}_{r=0}^{\mathcal{R}}$, to identify the transmitted symbol from the observation $\xi $, where $\mathcal{R}$ represents the number of transmitting vectors in the constellation set, we have
\begin{subequations}\label{eqn41}
	\begin{align}
		&d_r= \frac{1}{2} \left[\xi^{(r)} +\xi^{(r+1)}\right]  \tag{40a}, \\
		&\hat{\bm s}=\bm{s}_r \qquad \text{if} \quad d_{r - 1} \le \xi < {d_r}\tag{40b},
	\end{align}
\end{subequations}
where $ d_0 = -\infty$, and $ d_{\mathcal{R}} = +\infty$. Considering the statistical characteristics previously discussed, the probability that the receiver correctly identifies the transmitted signal $\bm{s}_r$  is given by 
\begin{equation}\label{eqn42}\tag{41}
\mathbb{P}(\bm{s}_r) = \int_{d_{r - 1}}^{d_r} f_{\xi} \left (x| \sigma_{\xi}^2,\bm{s}_r \right ) dx,	
\end{equation}  
where $f_{\xi}(x)$ is the pdf of $\xi $. According to the properties of the Gaussian distribution, the \eqref{eqn42} can be expressed as 
\begin{equation}\label{eqn43}\tag{42}
	\mathbb{P}(\bm{s}_r) = \Phi(d_r)-\Phi(d_{r - 1}).
\end{equation} 
Additionally, the error function can express this relationship, allowing the \eqref{eqn43} to be rewritten in a closed-form as 
\begin{equation}\label{eqn44}\tag{43}
\mathbb{P}(\bm{s}_r) =\frac{1}{2} erf \left(\frac{d_r-\mu_{\xi}}{\sqrt{2}\sigma_{\xi}^2} \right)-\frac{1}{2} erf\left(\frac{d_{r-1}-\mu_{\xi}}{\sqrt{2}\sigma_{\xi}^2} \right).
\end{equation}

Furthermore, the average symbol error rate (SER) at the receiver can be expressed as 
\begin{equation}\label{eqn044}\tag{44}
	\mathbb{P}_e =1-\frac{1}{\mathcal{R}} \sum \limits_{r = 1}^{\mathcal{R}} \mathbb{P}(\bm{s}_r). 
\end{equation}
\begin{figure}[!t]
	\centering
	\includegraphics[width=3.6in]{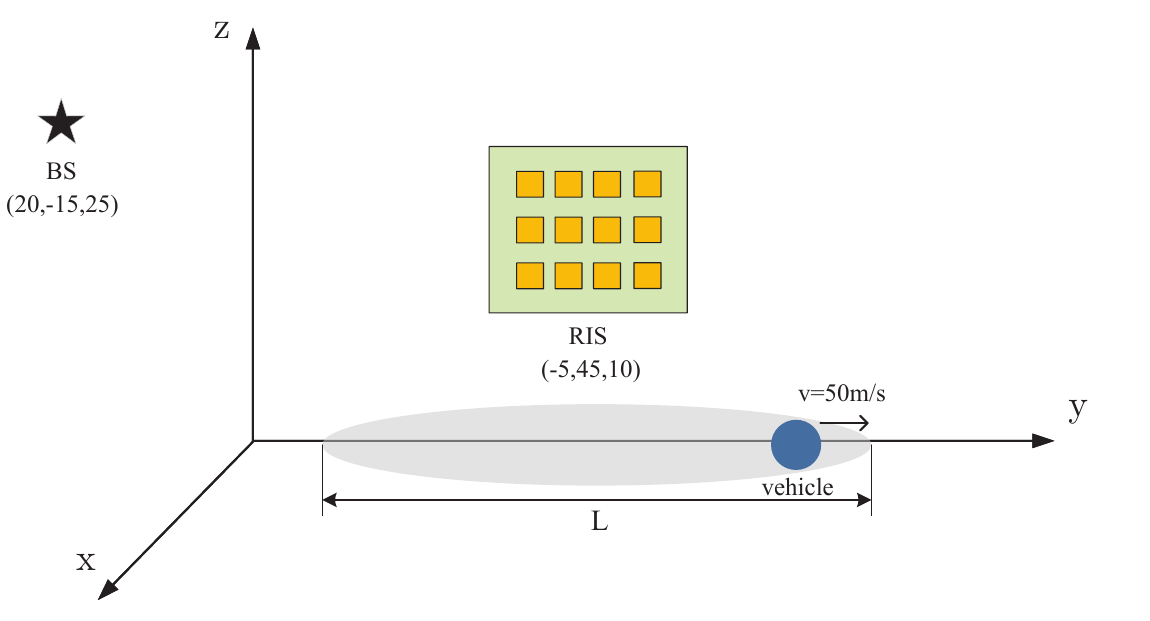}
	\caption{Simulation setup for roadside RIS-aided high-speed vehicle communication.}
	\label{Fig2}
\end{figure}
\section{Simulation Results}
\begin{figure*}[t]
	\centering
	\subfloat[]{
		\label{Fig3a}
		\includegraphics[width=0.5\textwidth]{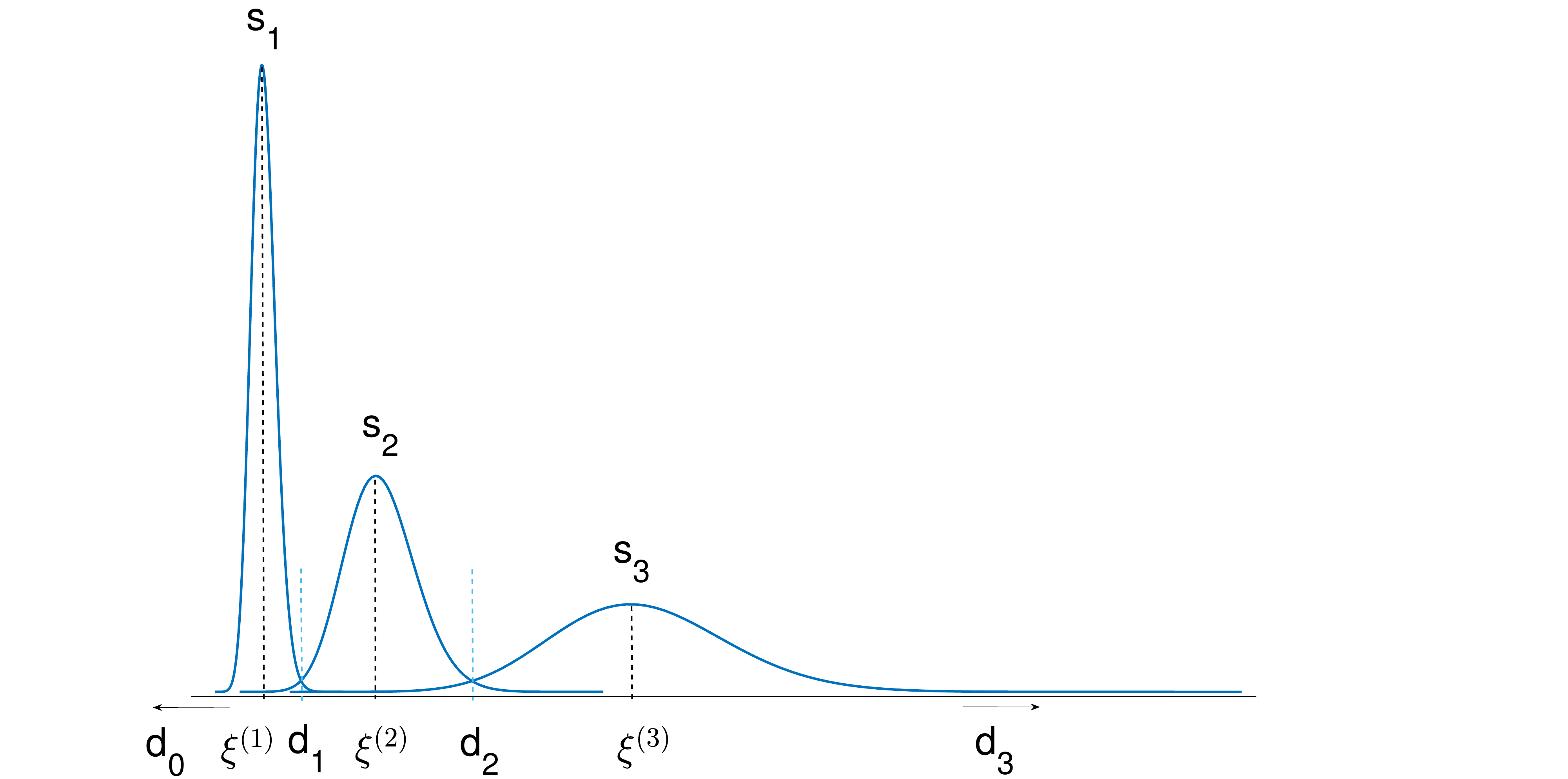}
	}
	\subfloat[]{
		\label{Fig3b}
		\includegraphics[width=0.5\textwidth]{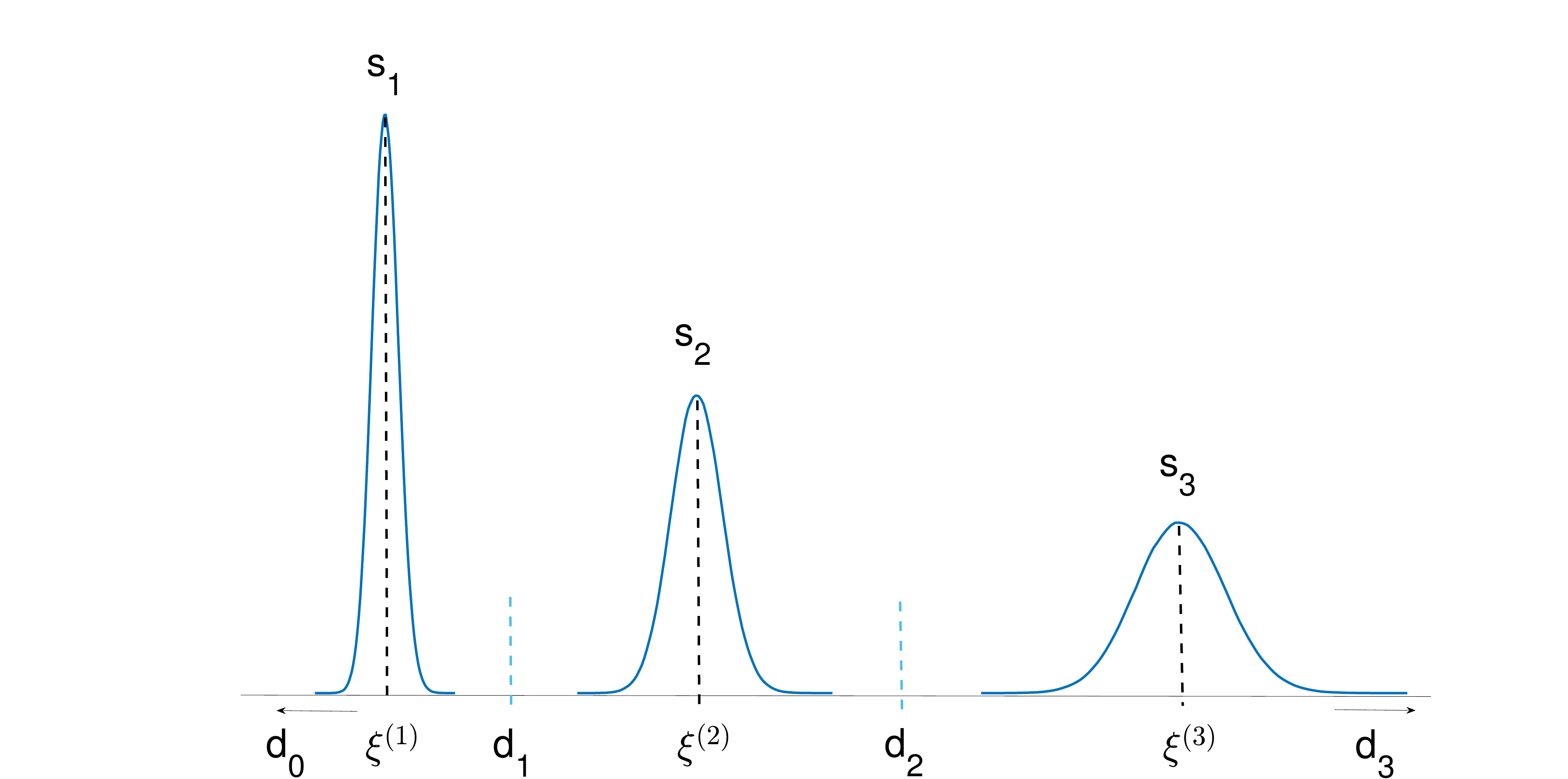}
	}
	\caption{Signal distribution in Euclidean norm detection with $s_1,s_2,s_3$, where (a) low SNR, (b) high SNR. }
	\label{Fig3}	
\end{figure*}
\begin{figure}[t]
	\centering
	\includegraphics[width=3.4in]{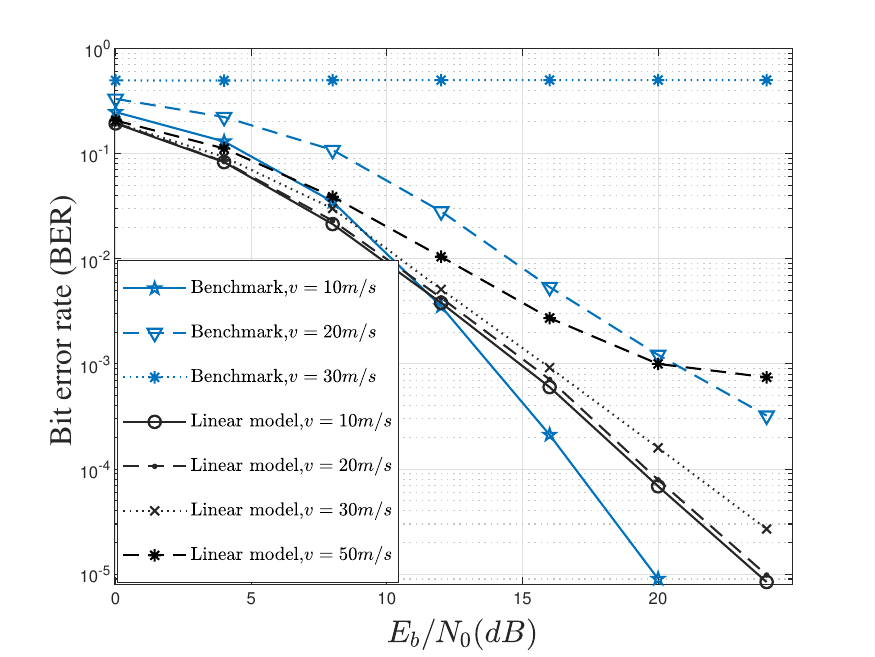}
	\caption{Robustness analysis of RIS-aided downlink communications at various mobile speeds. }
	\label{Fig4}
\end{figure}

In this section, the simulation results are presented to validate the performance of the proposed linear model for RIS-aided high-mobility communication systems. The initial setup of the RIS-aided vehicular communication system is shown in Fig. \ref{Fig2}. The positions of the BS and RIS are (20, -15, 25) meter (m) and (-5, 45, 10) m, respectively. Additionally, $L$ represents the RIS coverage distance, and the vehicle traverses this area while communicating with the BS, with $N_k=8, N=64$, and $N_t=128$. The link connecting the BS and RIS is characterized using the generalized clustered delay line approach outlined in 3GPP TR 38.901 Release 16, with the Rician factor set to $10$. Considering the different distances and LoS availability for the BS-user, BS-RIS, and RIS-user links, the path loss exponents are set to 2.5, 2.3, and 2.1, respectively\cite{ref40}. Meanwhile, the signal bandwidth is set to $2.5\;\rm{M H z}$ and each symbol duration is $T_s=8 \; \rm{\mu s}$. Each frame is composed of 40 blocks and contains 1,020 symbols, which includes a 20-bit Hadamard matrix used as the training sequence. To ensure the orthogonality of the two frequencies, we set the carrier frequencies to $f_{\rm 1}=5.9\;\rm{G H z}$, as specified by both the 3GPP and IEEE 802.11g standards for cellular V2X applications, and $f_{\rm 2}=f_{\rm 1}+ 1/T_s $. Given that the vehicle speed is $50 \;\rm{m/s}$, the maximum Doppler frequency is $ f_{\rm max}= vf_{\rm c}/\rm c\approx 1\;\rm {K H z} $, where $ c = 3\times 10^8 \;\rm{m/s}$ denotes the speed of light. The spacing both adjacent BS antennas and RIS elements is configured to half-wavelength distances. It is assumed that the AoA and AoD are uniformly distributed and randomly generated. Concurrently, the NLoS part of the channel is assumed to follow Rayleigh fading, with its time correlation modeled according to Jake's spectrum\cite{ref08}.
\begin{figure}[t]
	\centering
	\includegraphics[width=3.4in]{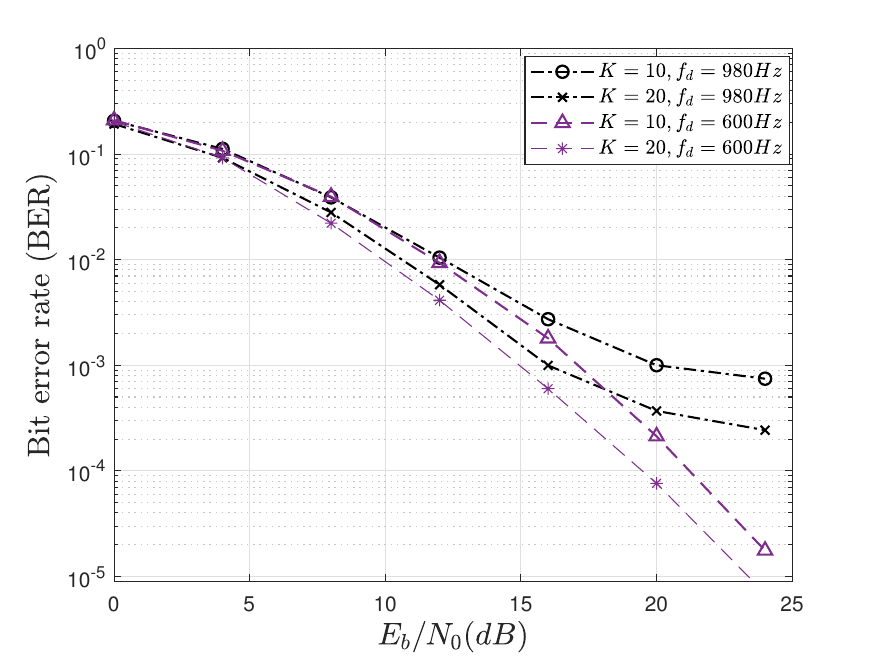}
	\caption{BER versus $E_b/N_0$ under various Rician factors and Doppler shifts.}
	\label{Fig5}
\end{figure}

In Fig. \ref{Fig3}, we illustrate the distribution curves of received signals under various SNRs when the receiver employs minimum Euclidean norm based on the linear model for signal detection. Notably, $\bm{s}_r$ denotes the transmitted signal, and $d_r$ represents the boundaries of different decoding regions. According to the decision criterion based on the signal's probability distribution, if $\xi$ is on the left side of the decision boundary $d_1$, it indicates that the transmitted signal is $\bm{s}_1$. If $\xi$ is between decision boundaries $d_1$ and $d_2$, it corresponds to the transmitted signal $\bm{s}_2$. Lastly, if $\xi$ falls to the right of decision boundary $d_2$, it signifies the transmitted signal $\bm{s}_3$. Therefore, it is evident from the Fig. \ref{Fig3} that different signal transmission qualities significantly affect the accuracy of signal detection and the delineation of signal decision region boundaries. The primary factors are outlined as follows. Under low SNR conditions, transmission signals exhibit higher variance, resulting in overlapping distributions of adjacent signals. Simultaneously, greater noise causes shifts in the centers of these distributions, further reducing detection accuracy. In contrast, under high SNR conditions, reduced variance leads to more concentrated signal distributions, decreasing overlap between adjacent signals and thereby improving detection performance. 

In the downlink fast-fading scenario, we compare the interference resilience of a linear model with a benchmark scheme. In addition, the benchmark scheme employs quadrature amplitude modulation (QAM) and ML detection, operating within the same channel environment as the linear model. It is observed from Fig. \ref{Fig4} that as the user's movement speed increases, the communication performance of the benchmark model gradually deteriorates, with speeds exceeding $30\;\rm m/s$ severely affecting communication capability. In contrast, the linear system consistently maintains relatively stable communication capability over a wide range of speed variations. However, in conditions of low Doppler shift, the performance of the linear model declines compared to the benchmark model. This phenomenon primarily arises from the linear model using an equivalent real-domain channel for signal detection to reduce the impact of rapid phase variation on communication. However, this method also results in the loss of channel phase information, thereby affecting system performance. In addition, when the user speed reaches $50\;\rm m/s$, the performance curve of the linear system gradually stabilizes with increasing SNR. It is suggested that further increases in SNR do not significantly improve the detection performance. This limitation is primarily due to large Doppler shifts, which introduce significant errors in channel estimation and increase interference between adjacent signals. Consequently, despite the higher SNR, the system fails to mitigate the adverse effects of Doppler shift, resulting in only limited improvement in system performance.
 	
In Fig. \ref{Fig5}, we compare the BER versus $E_b/N_0$ under various Rician factors and Doppler shifts. Several interesting observations are made as follows. First, it is observed that for the linear system, the system's robustness gradually improves as the Rician factor $K$ increases, under the same Doppler shift. This indicates that a strong LoS component significantly enhances the stability of information transmission in fast-fading environments. Additionally, the proposed signal design and detection algorithm are tailored for Rician channels dominated by the LoS component, with a higher Rician factor further enhancing the algorithm's effectiveness. Second, under the same Rician factor, system performance gradually declines as the Doppler shift increases. This can be explained by the fact that larger Doppler shifts reduce the channel coherence time, intensifying time-selective fading. Moreover, higher Doppler shift increases Doppler spread and amplifies system noise, further degrading the system's ability to suppress interference. Nevertheless, the Doppler-robust linear model is designed for Rician fading environments dominated by the LoS path. Increasing the Rician factor partially mitigates the effects of Doppler shifts, improving the stability of the communication system.

\begin{figure}[t]
	\centering
	\includegraphics[width=3.4in]{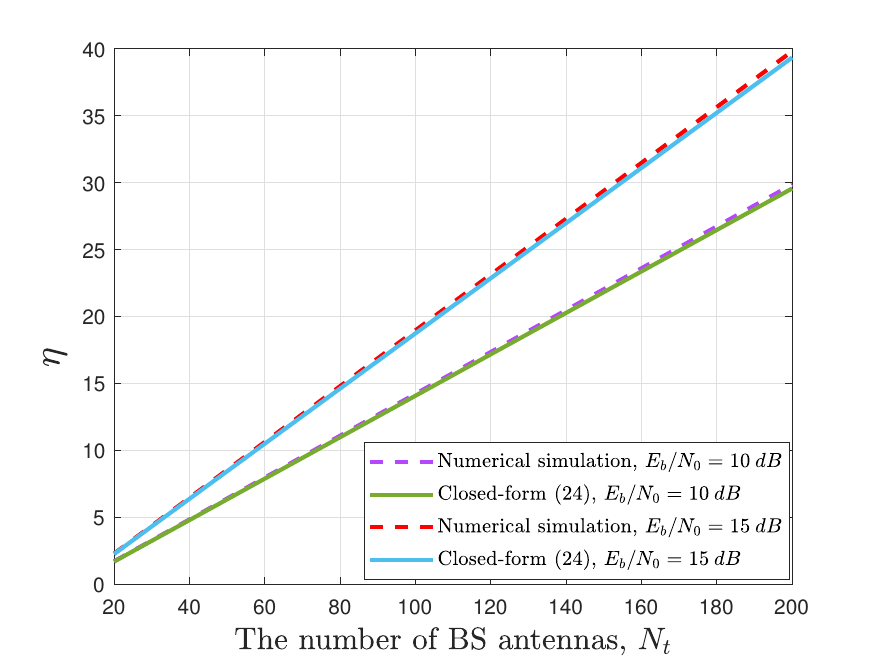}
	\caption{Output SNR of RIS-aided downlink precoding system.}
	\label{Fig6}
\end{figure}
\begin{figure*}[t]
	\centering
	\subfloat[]{
		\includegraphics[width=2.4in]{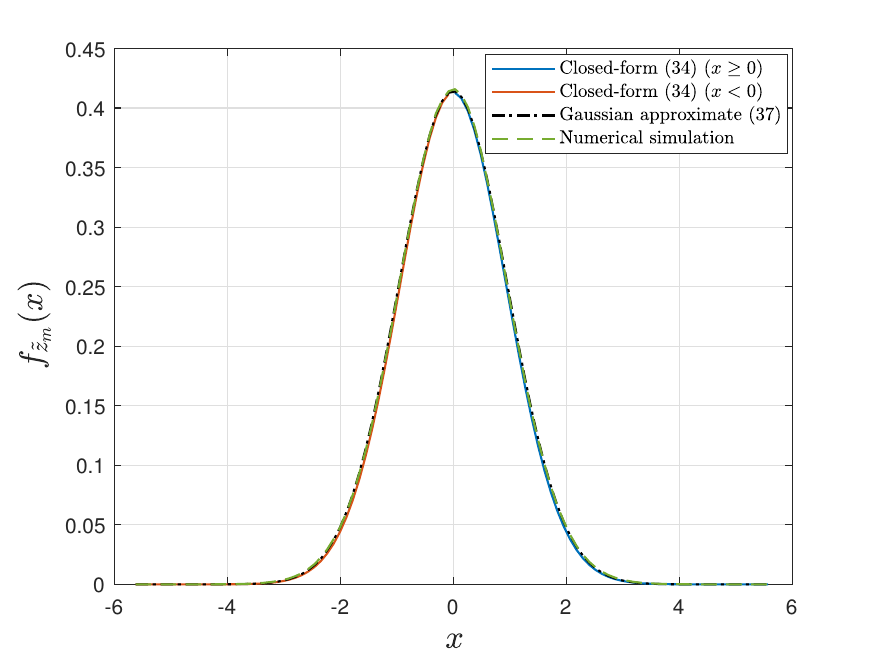}
		\label{Fig7a}
	}
	\subfloat[]{
		\includegraphics[width=2.4in]{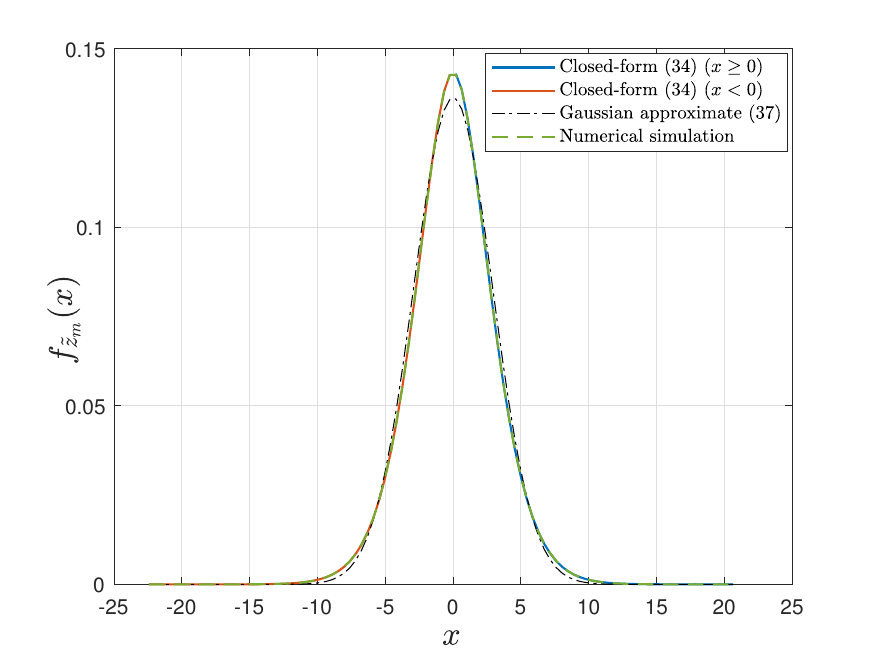}
		\label{Fig7b}
	}
	\subfloat[]{
		\includegraphics[width=2.4in]{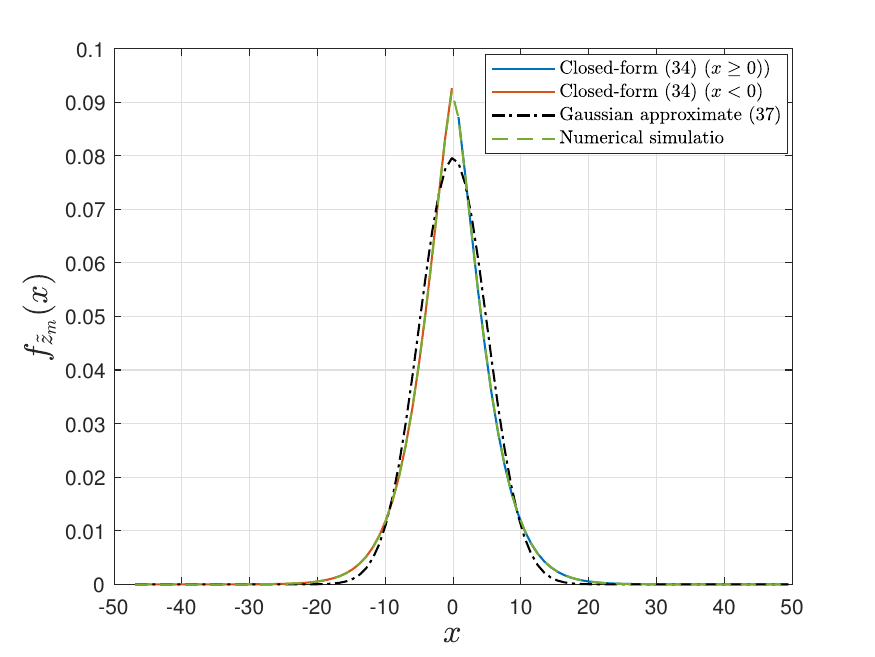}
		\label{Fig7c}
	}
	\caption {$f_{\tilde{z}_m} \left(x\right)$ of RIS-aided communications system at various SNR, where (a) high SNR, (b) medium SNR, (c) low SNR.}
	\label{Fig7}
\end{figure*}
\begin{figure}[t]
	\centering
	\includegraphics[width=3.4in]{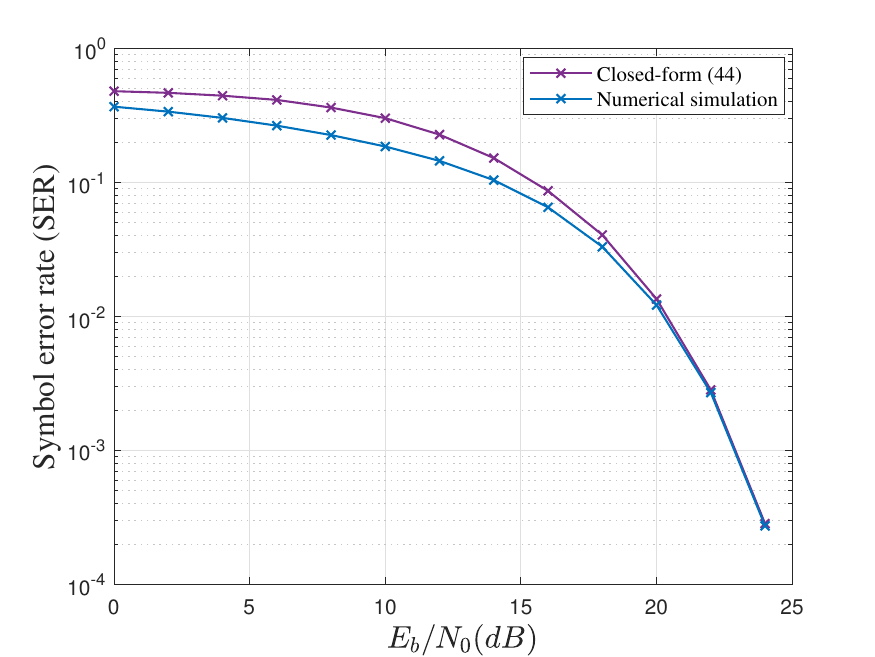}
	\caption{SER of RIS-aided uplink communications system.}
	\label{Fig8}
\end{figure}

Previous studies have analyzed the implementation process in precoding design based on the linear model, and derived theoretical expressions in closed-form for the output SNR of precoding systems. In Fig. \ref{Fig6}, we elaborate how the output SNR of the precoding system varies with the number of transmit antennas. Simulation results reveal a congruence between the theoretical formula and the simulation curve of the system's output SNR, thus validating the accuracy of the analysis. Moreover, the fit of the two curves remains unchanged regardless of the transmitter's SNR. It is clear that as the number of transmitting antennas increases, the impact of the transmitter's SNR on the receiver's output SNR becomes more noticeable. Notably, the output SNR increases linearly with the number of transmit antennas, indicating that adding more antennas at the BS can significantly enhance overall system performance. The specific explanation is that increasing the amount of transmitting antennas can enhance diversity gain, strengthen beamforming capability, and thereby improve the system's output SNR. 

In Section \uppercase\expandafter{\romannumeral4}, we utilize the characteristics of RIS-aided uplink communication systems to approximate the probability distribution of received signals using a Gaussian approximation, aiming to simplify the performance analysis of the system. In Fig. \ref{Fig7}, we compare the fitting between the probability distribution curve of the received signal and the Gaussian approximation under various SNRs. It can be observed that the received signal distribution curve, derived from \eqref{eqn35}, closely matches the numerical simulation results across different SNRs. Moreover, as the SNR decreases, the accuracy of the Gaussian approximation for the probability distribution of the received signal gradually declines. The main reason is that in the improved linear model, the probability distribution of the received signal follows a generalized gamma distribution, which approximates a Gaussian distribution only at high SNR\cite{ref38,ref39}. At the same time, the accuracy of the Gaussian approximation also impacts the performance analysis of the system. By approximating the distribution of the received signal with a Gaussian model, we simplify the Pdf, thereby reducing the complexity of the system performance analysis.

In the performance analysis of the improved linear model, some approximations are introduced to simplify the analysis, which can lead to inaccuracies. Therefore, it is necessary to compare the fitting of the average SER curve, derived from \eqref{eqn044}, with the numerical simulation results at different SNRs. As shown in Fig. \ref{Fig8}, as the SNR increases, the match of the asymptotic performance curve with the numerical simulation results improves. However, as the SNR decreases, the fitting between them gradually deteriorates. Specifically, the reason for this phenomenon is that the SER of the improved linear model is derived based on the Gaussian approximation results. It is observed from Fig. \ref{Fig7}, the differences in the fitting degrees of the Gaussian approximation and probability distribution curves at different SNRs lead to variations in the fitting degrees of the two curves in Fig. \ref{Fig8} as the SNR changes. Besides, even minor deviations in the match of the probability distribution curve can significantly affect the accuracy of the asymptotic performance analysis.
\begin{figure}[t]
	\centering
	\includegraphics[width=3.4in]{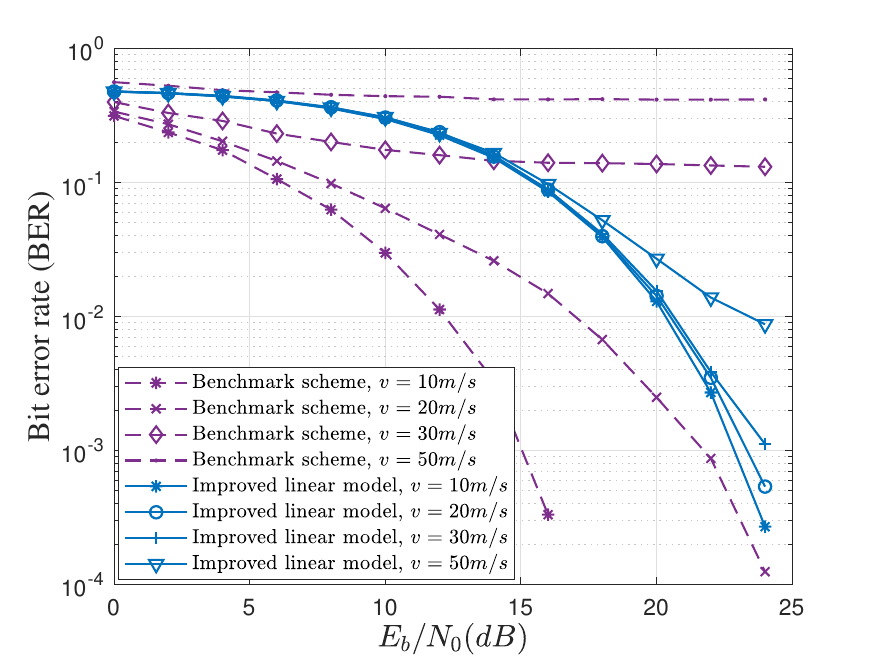}
	\caption{Robustness analysis of RIS-aided uplink communications at various mobile speeds.}
	\label{Fig9}
\end{figure}

In Fig. \ref{Fig9}, we plot the BER versus energy per bit to noise power spectral density ratio $(E_b/N_0)$ for the improved linear model and a benchmark scheme in the RIS-aided uplink fast-fading scenario. In the benchmark scheme, the transmitting end employs QAM modulation, while the receiving end uses ML detection, operating in the same channel environment as the improved linear model. From the Fig. \ref{Fig9}, it is evident that in the RIS-aided uplink communication system, as the user's movement speed increases, the communication performance of the benchmark scheme fails to meet the communication requirements. In contrast, the improved linear system maintains robust communication capabilities even under significant speed variations. Moreover, in the process of system modeling and signal detection, employing the law of large numbers and an equivalent real-domain linear model may result in reduced performance of the improved linear model compared to the benchmark scheme in low-speed scenarios. Specifically, the reason for this phenomenon is that the improved linear model eliminates the need for intricate channel estimation. By employing a real-domain linear model for signal detection, it not only simplifies channel estimation and synchronization but also enhances the system's Doppler robustness. However, as the velocity of movement increases, significant Doppler shifts notably influence the precision of channel estimation and introduce substantial inter-symbol interference. Consequently, even with enhanced SNR, the system's performance does not exhibit noticeable improvement. 

\section{Conclusion}
In this paper, we explored an innovative RIS-aided high-mobility communication system with the architecture of RIS-integrated BS, and proposed a linear model that ensures reliable  transmission and reduces channel estimation overhead. Specifically, we performed a series of signal processing techniques, including signal design, matched detection algorithms, and precoding, to improve the stability of high-mobility downlink communication systems. Subsequently, leveraging the reciprocity of uplink and downlink channels, along with the characteristics of the uplink communication system, we designed an improved uplink linear model based on the downlink linear model. We then analyzed the asymptotic performance of the improved linear model and derived closed-form expressions to gain further insights. Simulation results validated our analysis and demonstrated the performance advantages of the proposed RIS-aided high-mobility linear architecture over other benchmark schemes.

In the future, it will be an interesting direction to study more practical cases, including node discovery, beam alignment/tracking, coordinated handover, and efficient CSI acquisition, which will require more complicated system designs and hardware requirements. Moreover, extending the above designs to more promising scenarios, how to employ active RIS to achieve broader coverage, mount dedicated sensors (e.g., rotatable antenna (RA)-based sensors\cite{ref013}) on the RIS to enhance RIS's perceptive capabilities, and deploy STAR-RIS to enable more flexible coverage areas is also a challenging problem to solve.

\appendix[Proof of the generalized gamma distribution]
By denoting $\bar r=\bm {c}_m \bm{s}$ and $A=|\bm{c}_m \bm s + v_m^{(1)}|$, thus $\varepsilon_m$ can be rewritten as  
\begin{equation}\label{eqn45} \tag{45}
	\varepsilon_m= A^2,	A\geq 0,	
\end{equation} 
where we have
\begin{equation}\label{eqn46} \tag{46}
	A=\sqrt{\varepsilon_m},	\varepsilon_m\geq 0.	
\end{equation} 

Since $v_m^{(k)}$ is a zero-mean AWGN with variance $\sigma_v^2$, we have
\begin{equation}\label{eqn47} \tag{47}
	\bm{c}_m \bm s + v_m^{(1)}\sim \mathcal{CN} (\bar r,\sigma_v^2).	
\end{equation} 

It is apparent that the envelope $A$ follows a Rician distribution, we obtain the pdf of the  $A$ as
\begin{equation}\label{eqn48} \tag{48}
f_{A} (t)
=\frac{t}{\sigma_v^2} {\rm I}_{0} \left(\frac{\bar r t } {\sigma_v^2 } \right) \exp \left(- {\frac{\bar r^2+t^2}{2\sigma_v^2 }} \right), t\geq 0.	
\end{equation} 

From the \eqref{eqn45} and \eqref{eqn46}, as well as the functional distribution criterion for random variables, we can infer the pdf of the $\varepsilon_m$ as
\begin{equation}\label{eqn49} \tag{49}
	\begin{split}
		f_{\varepsilon_m} (x)&=f_{A}(t) \cdot \frac{d{A(x)}}{dx}  \\
		&= \frac{\sqrt{x}}{\sigma_v^2} {\rm I}_{0} \left(\frac{\bar r \sqrt{x} } {\sigma_v^2 } \right) \exp \left(- {\frac{\bar r^2+x}{2\sigma_v^2 }} \right)\frac{1}{2\sqrt{x}}\\
		&=\frac{1}{2\sigma_v^2}{\rm I}_{0} \left(\frac{\bar r \sqrt{x} } {\sigma_v^2 } \right) \exp \left(- {\frac{\bar r^2+x}{2\sigma_v^2 }} \right), x\geq 0.
	\end{split}   	
\end{equation}

Notice that the pdf of the generalized gamma distribution in \eqref{eqn33} can be directly obtian in \cite[(1)]{ref38}. By substituting $\alpha=1$, $\beta=2\sigma_v^2$, and $\gamma=\bar r^2$ into \eqref{eqn33}, we arrive at the expression given in \eqref{eqn49}, thus completing the proof.
\bibliographystyle{IEEEtran}
\bibliography{Reference}
\nocite{*}
\end{CJK}

 \end{document}